\documentclass[a4paper,10pt,titlepage]{article}
\usepackage{graphicx,epsfig}

\makeindex

\title{A brief overview on the BioPAX and SBML standards for formal
presentation of complex biological knowledge}
\author{Leo Lahti \\ University of Helsinki \\ leo.lahti@iki.fi}
\date{2007}

\begin{document}

\maketitle
\newpage

\noindent
Copyright \copyright 2007-2011 Leo Lahti\\
Some Rights Reserved.\\
http://www.iki.fi/Leo.Lahti (leo.lahti@iki.fi)\\[2mm]

\noindent
\includegraphics[width=4.5cm]{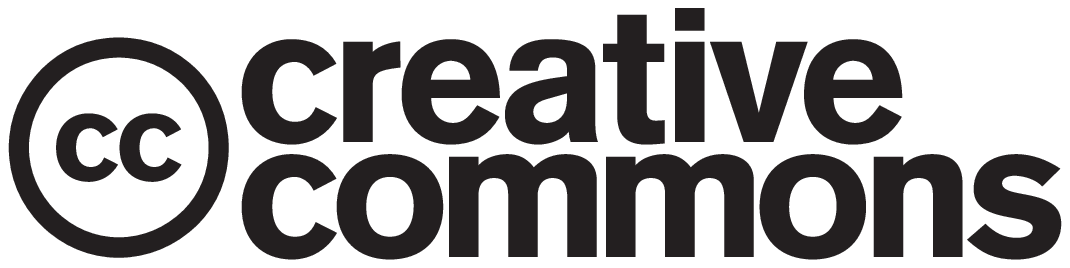}

\noindent This work is licensed under the terms of {\it Creative
	  Commons Attribution 3.0 Unported} license available from
	  http://www.creativecommons.org/. Accordingly, you are free
	  to copy, distribute, display, perform, remix, tweak, and
	  build upon this work even for commercial purposes, assuming
	  that you give the original author credit. See the licensing
	  terms for details.

	    	    
\newpage

\section{Background}

The quest for systems-level understanding of cellular mechanisms is a major para\-digm in modern biology \cite{Collins03}. Development of various high-throughput technologies for measuring cell-biological phenomena has accelerated systems biology research through rapid accumulation of biological data in public and private repositories.

Various databases focus on biological pathways, structured representations of cellular processes. Pathways are idealized descriptions of molecular events within the cell, and represent the current state of knowledge of these systems. Multiple proteins and other compounds are typically involved in a hierarchical manner through various molecular mechanisms, forming a complex biological network. Pathway data exhibits many details of the well-studied parts of the system, and can potentially be used to guide the search for novel cellular mechanisms when combined with high-throughput measurements such as gene expression, metabolomics or proteomics data (see e.g. \cite{Novak06}). Computational modeling can also highlight aspects of the existing networks that have been left unnoticed by more traditional methods.

Utilizing the huge data collections to create new biological knowledge is a challenge where novel computational tools play a key role \cite{Cohen04}. To take full advantage of the possibilities opened up by large-scale community databases, efficient standards for representing biological information are needed. This will enhance the sharing and evaluation of knowledge, and development computational tools. 

A good standard is independent of the software environment where the data is created and analyzed, and ensures the usability of the data beyond the lifetime of any single software. Perhaps surprisingly, the standards for representing biological network data have emerged only recently. In the beginning of 2000's, the field was lacking proper standards, and the integration of network data from different databases required considerable effort as each database was using its own internal representation and analysis tools. 

Various frameworks for the representation of network information were proposed to alleviate these problems at the time. In this work, I present an overview of two such proposals, {\it BioPAX} and {\it SBML}. These two approaches have gained popularity and are promising candidates to become established as widely adopted future standards in the field.

BioPAX and SBML are advancing the construction of databases that can be widely used to model, simulate, and visualize network data in various programming languages. They are not competing standards but rather complementary approaches to describe related phenomena. SBML is targeted at modeling systems of dynamic biochemical reaction networks that are described by reaction equations. BioPAX assumes a static view on the network model. It does not contain mathematical formulas but provides more detailed information concerning the individual molecules and interactions.

This work is mainly based on the technical manuals of BioPAX (2.1) and SBML (2.3 prerelease), their respective websites\footnote{http://www.biopax.org, http://sbml.org}, and the comparative review of these representation schemes by Str\"omb\"ack and Lambrix \cite{Stromback05}. The work is organized as follows. First, I will illustrate the nature of biological network data through a simple example. Then BioPAX and SBML are considered in more detail. The general scope and the overall structure of both standards is described. Finally, various aspects of the two approaches, and the differences between them are discussed. 

The aim of this work is to give a good overall picture of these standards. Hence I have skipped many details of the two models in this short introduction.

\section{Biological network models}

Biological network models describe complex molecular systems that are active in certain biological conditions. Knowledge of these systems typically arises from various individual studies on different parts of the network. Traditionally, biochemical networks have been summarized in schematic two-dimensional maps. Until recently, textbook figures of such maps have been the only general description of the system for many networks. For large-scale computational studies, such illustrations are not sufficient, and the need for computer-readable representations is evident.

\subsection{A simple example of a biological network model}

\begin{figure}[h]
\epsfig{file=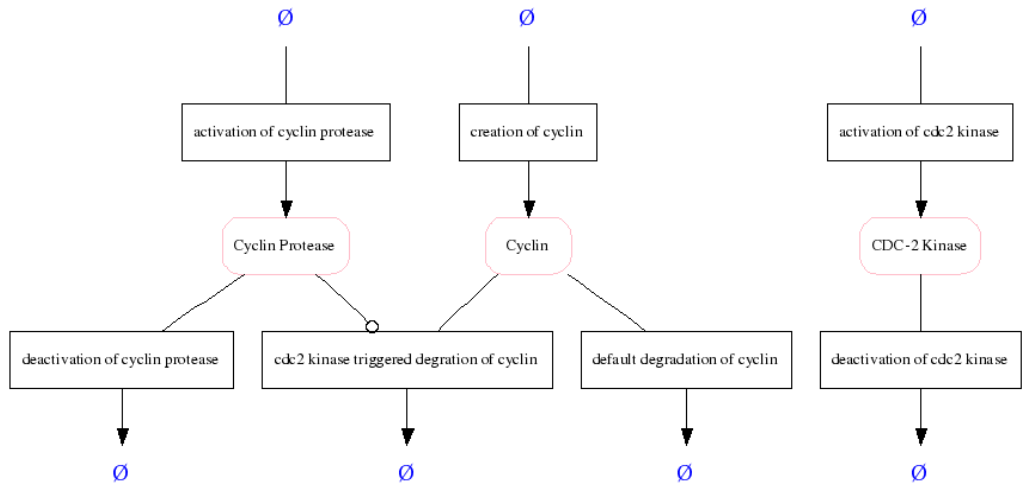}
\caption{Visualization of the network model BIOMD0000000003 from the BioModels database.}
\label{fig:BioModel}
\end{figure}

The BioModels database\footnote{http://www.ebi.ac.uk/biomodels/} is one of the several public databases that currently provide information on biological networks. The {\it minimal cascade model for the mitotic oscillator involving cyclin and cdc2 kinase} \cite{Goldbeter91} picked from this database is a simple example of a molecular network model\footnote{http://www.ebi.ac.uk/compneur-srv/biomodels-main/publ-model.do?mid=BIOMD0000000003} (Figure~\ref{fig:BioModel}).

In this example, as in other typical illustrations of a network model, nodes represent the involved biomolecules, and edges indicate the interactions between these molecules. A database may contain additional information concerning the individual molecules and the nature of the interactions, although such information may not be visualized in simple interaction maps. A complex dynamical system, for example, cannot be illustrated within one static figure, whereas a carefully designed database structure can make this possible. Hyperlinking within and between databases is also made possible by current database technologies, and allows easy navigation of the network data for application-oriented users.

The network models in the BioModels database can be exported in both BioPAX and SBML format. These describe the same network but provide complementary information of it. The XML-based implementation of both models are given for the cascade network of Figure~\ref{fig:BioModel} in the Appendix. 

Even for describing such a simple network, somewhat lengthy code is apparently needed to capture any details of the model. The code is here offered for demonstration purposes and is not considered in detail; the intention of this work is to provide a general overview of the two network modeling standards- not to single out details of a specific biological network. The general characteristics of the BioPAX and SBML schemes, and their differences are detailed below. 

\section{Standards for biological network information}

\subsection{Biological Pathways Exchange Language (BioPAX 2)}

BioPAX \cite{BioPAX05} is a specific data type for representing and sharing information concerning biomolecular networks\footnote{http://www.biopax.org}. BioPAX does not present mathematical relations, but provides many details concerning the molecular interactions within the cell. The current release of BioPAX (Level 2) can be used to represent metabolic networks and molecular interaction networks. Support for gene and DNA interactions, genetic interactions, and signal transduction are planned for upcoming releases.

A number of popular network databases support the BioPAX format,  including BioCyc\footnote{http://biocyc.org}, Reactome\footnote{http://reactome.org/}, BioModels Database, and Pathway Interaction Database\footnote{http://pid.nci.nih.gov}. Various network analysis tools for BioPAX are available as well\footnote{http://biopaxwiki.org/cgi-bin/moin.cgi/Biological\_software\_supporting\_BioPAX}, many of them open source. One of the most popular tools for viewing and editing BioPAX ontologies is Protege\footnote{http://protege.stanford.edu}, a user-friendly software with a graphical user interface. For more computationally oriented people, some tools for low-level analysis of the BioPAX ontology are offered by the Rredland package\footnote{http://bioconductor.org/packages/2.0/bioc/html/Rredland.html} of the R/BioConductor project, for example. It is worth emphasizing that the scope of these software environments is not restricted to any single database but can handle data from any database supporting the BioPAX format.

\subsubsection{A brief characterization of the BioPAX network model}

BioPAX describes its objects in an inheritance class hierarchy. The basic building blocks of a BioPAX model are the {\it Physical Entity}, {\it Interaction} and {\it Pathway} classes. {\it Pathway} represents a set of interactions that together form a network model. Each {\it Interaction} describes relationships between {\it Physical Entities}. This structure is the core of a BioPAX network model. 

Details concerning the interactions and entities are given by subclasses that describe various interaction and physical entity types. Physical entities are various objects that may interact, including molecular complexes, proteins, DNA, RNA, and small molecules. Each interaction type limits the possible roles and the number of potentially interacting entities. Transport, catalysis and modulation are examples of possible interactions. 

In summary, BioPAX specifies the type of the interacting molecules and details of their interactions in a pre-defined class hierarchy. Such representation enables efficient sharing of the information between databases. To obtain sufficient generality in the model for such purposes, many details have to be omitted in the model hierarchy. However, additional information and hyperlinks to external sources can be assigned to BioPAX objects. It is also possible to make application-specific additions to the BioPAX standard model, although this is not recommended.

BioPAX code for the network model in Figure~\ref{fig:BioModel} is given in the Appendix. The source code contains all information encoded into the model but it is not meant to be human-readable as such. Appropriate visualization and analysis tools help in the interpretation of the data. The example code illustrates the internal structure of a BioPAX model and is self-explanatory for people with previous experience on XML-based database programming.

\subsubsection{Rredland tool for R/BioConductor}

I briefly tested the R/BioConductor tool for BioPAX (Rredland) to see if I could utilize it in my own bioinformatics research. Unfortunately, this specific package still seems to be in an initial stage, and only simple queries are possible. However, the package is under active development and the release of a sufficient collection of low-level analysis routines can be expected soon. Currently, methods for reading the data and doing basic queries are available.

Following is a very simple example of BioPAX hierarchy for an interaction event in our example model in Figure~\ref{fig:BioModel}. I picked the 'creation of cyclin' event from the model. This appears to have the following attributes: 

\begin{tabular}{ll}
type 		&conversion\\
NAME 		&creation of cyclin\\
RIGHT	 	&reaction1\_RIGHT\_C
\end{tabular}

Accordingly, 'creation of cyclin' is a 'conversion' reaction, and this reaction has a relationship 'RIGHT' to another subject, 'reaction1\_RIGHT\_C'. Now we can proceed in inspecting the model and check the 'reaction1\_RIGHT\_C' subject in more detail. It appears to have the following attributes:

\begin{tabular}{ll}
type			&physicalEntityParticipant\\
PHYSICAL-ENTITY		&C\\
CELLULAR-LOCATION	&cell
\end{tabular}

Now we see that the subject 'reaction1\_RIGHT\_C' is of the type 'physicalEntityParticipant', the corresponding 'PHYSICAL-ENTITY' being 'C'. Also cellular location 'cell' is given although it is in this case very general type and hence rather uninformative. Further checking the attributes for 'C', we find that it is of 'physicalEntity' type with the name 'Cyclin'. 

In summary, 'creation of cyclin' is a conversion reaction that involves 'Cyclin' as its 'RIGHT' participant, denoting that Cyclin is the end product from this reaction. In the contrary, 'LEFT' would denote the input molecules for the reaction. For this specific event, additional details have not been stored in the model. 

This short query illustrates the basic structure of the BioPAX hierarchy from low-level analysis perspective. A typical user does not need to consider such raw representation schemes; application-oriented users can browse the network and related information with simple visualization tools. Further properties of our example network can be retrieved from the source code in the Appendix.

\subsection{Systems Biology Markup Language (SBML 2)}

SBML is another alternative to represent biological network information. The current version of SBML \cite{Hucka03} (level 2) is targeted at analysing and simulating basic biochemical reaction networks\footnote{http://sbml.org}. Examples of such networks include cell signaling, metabolic networks and gene regulation. SBML can be used to describe both quantitative and qualitative aspects of the reactions that modify or transport the entities. 

The SBML standard is supported by many popular network databases including KEGG\footnote{http://www.genome.jp/kegg}, Reactome, and the BioModels Database. Analysis of SBML data is supported by more than 110 software systems. Computational tools are available in several languages, including C++, MATLAB, Java and R/BioConductor\footnote{http://cran.r-project.org/src/contrib/Descriptions/rsbml.html}. 

In addition to the widely adopted XML format, the implementation of SBML model utilizes certain other standards, including MathML\footnote{http://www.w3.org/TR/2003/REC-MathML2-20031021} and the CellML\footnote{http://www.cellml.org}. In fact, the SBML standard is similar to the independent representation scheme CellML \cite{Lloyd04}. According to the SBML website, the editors of SBML and CellML have collaborated to fuse the two models to create a single standard in the future. 

\subsubsection{A brief characterization of the SBML model}

An SBML model consists of a set of {\it chemical species} that are linked by {\it reactions}. Chemical species include simple ions, simple molecules  (e.g. glucose), and large molecules such as RNAs or proteins. Each species is located in a {\it compartment} which is describes the reaction environment. The reactions can transform one entity into another, transport entities between compartments, or describe the binding of the species.
This is the general structure of an SBML model.

A number of more detailed features can be assigned to the objects of an SBML model. Chemical species can be assigned a spatial size and charge. In a given interaction, the participating species have particular roles such as {\it reactant}, {\it product} or {\it modifier}. It is also possible to specify whether a reaction is reversible. Mathematical description of the reactions are used to describe the reaction dynamics. The overall model can be provided with initial parameters, including the concentrations of the original species and constraints for the model parameters. Interestingly, SBML allows modeling of sudden discrete changes in the model under specified circumstances. An SBML model can even represent interactions between the interaction networks and other phenomena. 

SBML does not impose a class hierarchy into the model as in BioPAX. Instead, it focuses on mathematical description of the interactions. In general, SBML is not designed to store all potentially conflicting aspects of the network but to encode a coherent view of a biological reaction network that can be used in computational simulations. 

References to external information sources can be given in an annotation field. SBML also provides limited options for adding data that does not fit into the standard format. 

The SBML source code for the network model in Figure~\ref{fig:BioModel} is given in the Appendix for demonstration purposes, and can be compared with the BioPAX code describing the same network. Based on the documentation, the R package for analyzing SBML (RSBML) seems to be more complete than the R-based analysis tools for BioPAX. I personally need a general static description of the biomolecule relationships that I could combine with other high-throughput data such as gene expression measurements, for example. SBML is focused on dynamic simulations, and the biomolecular relationships in SBML are generally too complex for my current purposes. Therefore, I have not tested the RSBML package.

\section{Discussion}

Modern high-throughput technologies and standardized databases advance the accumulation of biological knowledge. This knowledge is often highly structured. Biological network data exemplifies the complex nature of biological data. Representation of dynamic biochemical reactions and the relationships between various compounds and molecular complexes in its full richness is a challenging task beyond a mere listing of events.

BioPAX and SBML are general frameworks to represent network information. These standards can be used to present complex systems-level biological knowledge that extends beyond listing sets of individual interactions between biomolecules, and allow the automated analysis of such data. This is different from standards like PSI-MI \cite{Hermjakob04} that offer detailed information on individual interactions but cannot be used to describe the more general system arising from these interactions. Various protein-protein databases that offer pairwise interaction data between biomolecules are reviewed in \cite{Mathivanan06}. Although SBML and BioPAX could be used for similar purposes, their focus is on describing a larger system of molecular mechanisms, and less attention is paid to individual interactions. If necessary, schematic pathway maps representing the pairwise interactions of a molecular network can be automatically generated from the network information provided by SBML or BioPAX.

BioPAX seems to combine the general representation of the network with sufficient details of the network objects better than SBML. Due to the hierarchical nature of BioPAX, a detailed representation of interactions and biomolecules can be obtained while allowing queries at a more general level. For example, the substructure of molecular complexes can be described in BioPAX while such option is only planned for the next level of SBML. 

The networks in BioPAX are static, and no information concerning reaction dynamics within the network is provided. On the contrary, SBML supports quantitative models and is a better format for dynamical simulations of biochemical networks. The increased mathematical complexity of SBML is balanced by a simplified representation of individual molecules and interactions. While BioPAX utilizes hierarchical representation to encode the the relationships between molecules in a biological network, SBML uses mathematical equations to define these relationships. SBML models are less structured than BioPAX models, but they give more quantitative details concerning the interactions and the biochemical reactions arising from these interactions. 

The utility of specialized standards depends on the fact that they restrict the representation to commonly agreed schemes. User-specific definitions and structures can cause problems for others handling the data, and do not fit in the idea of standardization. However, the possibility of including additional information is often essential in biomedical studies. Both standards offer limited possibilities for application-specific model extension. They also recommend that the standard names of biomolecules are used in the representations, but neither of them enforces the use of standard names.

The implementation of both BioPAX and SBML is based on XML, a widely used text-based format for representing hierarchical information schemes. The advantage of well-established ontology semantics is that it these models can be easily published in electronic form as supplementary data for peer-reviewed journal articles, for example. The use of standardized representations for systems-level biological knowledge will certainly advance the utilization of this knowledge in large-scale computational studies where the availability of such data in a suitable format is often a bottleneck. 

Both BioPAX and SBML have gained popularity in the bioinformatics community in the recent years, and are promising standards for sharing biological network information. The development of BioPAX and SBML is community-based. New features are added according to the needs of the bioinformatics community that uses these representation schemes. We are likely to see remarkable development and extensions in the two standards and related analysis tools as the user base grows and more people are contributing to the testing and development of these standards in one way or other. 

As the two standards are complementary rather than competing, it is not expected that one of them would override the other in the near future. However, as BioPAX and SBML are used to describe closely related aspects of biological networks, their future levels are likely to become more closely related. Currently, BioPAX offers a more detailed description of static molecular relationships within the network whereas SBML is better suited for computational simulations of network dynamics.


\begin{thebibliography}{10}

\bibitem{Collins03}
Francis~S. Collins, Eric~D. Green, Alan~E. Guttmacher, and Mark~S. Guyer.
\newblock A vision for the future of genomics research.
\newblock {\em Nature}, 422:835--847, 2003.

\bibitem{Novak06}
Barbara~A. Novak and Ajay~N. Jain.
\newblock {Pathway recognition and augmentation by computational analysis of
  microarray expression data}.
\newblock {\em Bioinformatics}, 22:233--241, 2006.

\bibitem{Cohen04}
Joel~E. Cohen.
\newblock Mathematics is biology's next microscope, only better; biology is
  mathematics' next physics, only better.
\newblock {\em PLoS Biology}, 2:e439, 2004.

\bibitem{Stromback05}
Lena Str\"omb\"ack and Patrick Lambrix.
\newblock {Representations of molecular pathways: an evaluation of SBML, PSI MI
  and BioPAX}.
\newblock {\em Bioinformatics}, 21:4401--4407, 2005.

\bibitem{Goldbeter91}
A.~Goldbeter.
\newblock A minimal cascade model for the mitotic oscillator involving cyclin
  and cdc2 kinase.
\newblock {\em {Proceedings of the National Academy of Sciences, USA}},
  88:9107--9111, 1991.

\bibitem{BioPAX05}
{BioPAX workgroup}.
\newblock {\em BioPAX - Biological Pathways Exchange Language}, 2005.
\newblock Level 2, Version 1.0 Documentation.

\bibitem{Hucka03}
M.~Hucka, A.~Finney, H.M. Sauro, H.~Bolouri, J.C. Doyle, {Kitano, H. and the
  rest of the SBML Forum}, A.P. Arkin, B.~J. Bornstein, D.~Bray,
  A.~Cornish-Bowden, A.A. Cuellar, S.~Dronov, E.~D. Gilles, M.~Ginkel, V.~Gor,
  I.~I. Goryanin, W.~J. Hedley, T.~C. Hodgman, J.-H. Hofmeyr, P.~J. Hunter,
  N.~S. Juty, J.~L. Kasberger, A.~Kremling, U.~Kummer, N.~Le~Novere, L.~M.
  Loew, D.~Lucio, P.~Mendes, E.~Minch, E.~D. Mjolsness, Y.~Nakayama, M.~R.
  Nelson, P.~F. Nielsen, T.~Sakurada, J.~C. Schaff, B.~E. Shapiro, T.S.
  Shimizu, H.D. Spence, J.~Stelling, K.~Takahashi, M.~Tomita, J.~Wagner, and
  J.~Wang.
\newblock {The systems biology markup language (SBML): a medium for
  representation and exchange of biochemical network models}.
\newblock {\em Bioinformatics}, 19(4):524--531, 2003.

\bibitem{Lloyd04}
Catherine~M. Lloyd, Matt D.~B. Halstead, and Poul~F. Nielsen.
\newblock {CellML: its future, present and past}.
\newblock {\em Progress in Biophysics and Molecular Biology}, 85:433--450,
  2004.

\bibitem{Hermjakob04}
H.~Hermjakob, L.~Montecchi-Palazzi, G.~Bader, R.~Wojcik, L.~Salwinski, A.~Ceol,
  S.~Moore, S.~Orchard, U.~Sarkans, C.~{von Mering}, B.~Roechert, S.~Poux,
  E.~Jung, H.~Mersch, P.~Kersey, M.~Lappe, Y.X. Li, R.~Zeng, D.~Rana,
  M.~Nikolski, H.~Husi, C.~Brun, K.~Shanker, S.G.N. Grant, C.~Sander, P.~Bork,
  W.M. Zhu, A.~Pandey, A.~Brazma, B.~Jacq, M.~Vidal, D.~Sherman, P.~Legrain,
  G.~Cesareni, L.~Xenarios, D.~Eisenberg, B.~Steipe, C.~Hogue, and R.~Apweiler.
\newblock The {HUPOPSI's} molecular interaction format - a community standard
  for the representation of protein interaction data.
\newblock {\em Nature Biotechnology}, 22:177--183, 2004.

\bibitem{Mathivanan06}
Suresh Mathivanan, Balamurugan Periaswamy, TKB Gandhi, Kumaran Kandasamy,
  Shubha Suresh, Riaz Mohmood, YL~Ramachandra, and Akhilesh Pandey.
\newblock An evaluation of human protein-protein interaction data in the public
  domain.
\newblock {\em BMC Bioinformatics}, 7(S5):S19, 2006.

\end{thebibliography}

\newpage

\section*{Appendix}

\subsection*{BioPAX}

BioPAX source code for the network model in Figure~\ref{fig:BioModel}.

\scriptsize
\begin{verbatim}

<rdf:RDF xml:base="http://www.ebi.ac.uk/biomodels/biopax">-<owl:Ontology rdf:about="">
<owl:imports rdf:resource="http://www.biopax.org/release/biopax-level2.owl"/>
</owl:Ontology>
-<bp:openControlledVocabulary rdf:ID="cell">
<bp:TERM>cell</bp:TERM></bp:openControlledVocabulary>-<bp:physicalEntity rdf:ID="C">
<bp:NAME>Cyclin</bp:NAME></bp:physicalEntity>-<bp:physicalEntity rdf:ID="M">
<bp:NAME>CDC-2 Kinase</bp:NAME></bp:physicalEntity>
-<bp:physicalEntity rdf:ID="X">
<bp:NAME>Cyclin Protease</bp:NAME></bp:physicalEntity>
-<bp:conversion rdf:ID="conversion_reaction1">
<bp:NAME>creation of cyclin</bp:NAME>
-<bp:RIGHT>-<bp:physicalEntityParticipant rdf:ID="reaction1_RIGHT_C">
<bp:PHYSICAL-ENTITY rdf:resource="#C"/><bp:CELLULAR-LOCATION rdf:resource="#cell"/>
</bp:physicalEntityParticipant></bp:RIGHT>
</bp:conversion>-<bp:conversion rdf:ID="conversion_reaction2">
<bp:NAME>default degradation of cyclin</bp:NAME>
-<bp:LEFT>-<bp:physicalEntityParticipant rdf:ID="reaction2_LEFT_C">
<bp:PHYSICAL-ENTITY rdf:resource="#C"/><bp:CELLULAR-LOCATION rdf:resource="#cell"/>
</bp:physicalEntityParticipant></bp:LEFT>
</bp:conversion>-<bp:conversion rdf:ID="conversion_reaction3">
<bp:NAME>cdc2 kinase triggered degration of cyclin</bp:NAME>
-<bp:LEFT>-<bp:physicalEntityParticipant rdf:ID="reaction3_LEFT_C">
<bp:PHYSICAL-ENTITY rdf:resource="#C"/><bp:CELLULAR-LOCATION rdf:resource="#cell"/>
</bp:physicalEntityParticipant></bp:LEFT>
</bp:conversion>-<bp:control rdf:ID="control_reaction3">
-<bp:CONTROLLER>-<bp:physicalEntityParticipant rdf:ID="reaction3_CONTROLLER_X">
<bp:PHYSICAL-ENTITY rdf:resource="#X"/><bp:CELLULAR-LOCATION rdf:resource="#cell"/>
</bp:physicalEntityParticipant></bp:CONTROLLER>
<bp:CONTROLLED rdf:resource="#conversion_reaction3"/>
</bp:control>-<bp:conversion rdf:ID="conversion_reaction4">
<bp:NAME>activation of cdc2 kinase</bp:NAME>
-<bp:RIGHT>-<bp:physicalEntityParticipant rdf:ID="reaction4_RIGHT_M">
<bp:PHYSICAL-ENTITY rdf:resource="#M"/><bp:CELLULAR-LOCATION rdf:resource="#cell"/>
</bp:physicalEntityParticipant></bp:RIGHT>
</bp:conversion>-<bp:conversion rdf:ID="conversion_reaction5">
<bp:NAME>deactivation of cdc2 kinase</bp:NAME>
-<bp:LEFT>-<bp:physicalEntityParticipant rdf:ID="reaction5_LEFT_M">
<bp:PHYSICAL-ENTITY rdf:resource="#M"/><bp:CELLULAR-LOCATION rdf:resource="#cell"/>
</bp:physicalEntityParticipant></bp:LEFT>
</bp:conversion>-<bp:conversion rdf:ID="conversion_reaction6">
<bp:NAME>activation of cyclin protease</bp:NAME>-<bp:RIGHT>
-<bp:physicalEntityParticipant rdf:ID="reaction6_RIGHT_X">
<bp:PHYSICAL-ENTITY rdf:resource="#X"/><bp:CELLULAR-LOCATION rdf:resource="#cell"/>
</bp:physicalEntityParticipant></bp:RIGHT>
</bp:conversion>-<bp:conversion rdf:ID="conversion_reaction7">
<bp:NAME>deactivation of cyclin protease</bp:NAME>
-<bp:LEFT>-<bp:physicalEntityParticipant rdf:ID="reaction7_LEFT_X">
<bp:PHYSICAL-ENTITY rdf:resource="#X"/><bp:CELLULAR-LOCATION rdf:resource="#cell"/>
</bp:physicalEntityParticipant></bp:LEFT></bp:conversion></rdf:RDF>

\end{verbatim}

\pagebreak

\subsection*{SBML}

\normalsize{SBML source code for the network model in Figure~\ref{fig:BioModel}.}

\scriptsize

\begin{verbatim}
<?xml version="1.0" encoding="UTF-8"?>
<sbml xmlns="http://www.sbml.org/sbml/level2" metaid="_180324" level="2" version="1">
  <model metaid="_180340" id="GMO" name="Goldbeter1991_MinMitOscil">
    <notes><body xmlns="http://www.w3.org/1999/xhtml">
    <p><h2><center>A Simple Mitotic Oscillator</center></h2></p>
    <p>Reference:Goldbeter A (1991)<i>A minimal cascade model for the mitotic oscillator involving cyclin 
	and cdc2 kinase</i>, PNAS 88:9107-9111<br></br>Web Reference:
	<a href="http://www.pnas.org/cgi/content/abstract/88/20/9107">
	http://www.pnas.org/cgi/content/abstract/88/20/9107</a></p>
    <p style="font-size:x-small;">This is a Systems Biology Markup Language (SBML) file, generated by 
	MathSBML 2.4.6 (14-January-2005) 14-January-2005 18:33:39.806932. SBML is a form of XML, and most 
	XML files will not display properly in an internet browser. To view the contents of an XML file use 
	the "Page Source" or equivalent button on you browser.</p>
   <p>This model originates from BioModels Database: A Database of Annotated Published Models. 
	It is copyright (c) 2005-2006 The BioModels Team.<br/> For more information see the 
	<a href="http://www.ebi.ac.uk/biomodels/legal.html">terms of use</a>.</p></body></notes>
<annotation><rdf:RDF xmlns:rdf="http://www.w3.org/1999/02/22-rdf-syntax-ns#" 
xmlns:dc="http://purl.org/dc/elements/1.1/" 
xmlns:vCard="http://www.w3.org/2001/vcard-rdf/3.0#" 
xmlns:dcterms="http://purl.org/dc/terms/" xmlns:bqbiol="http://biomodels.net/biology-qualifiers/" 
	xmlns:bqmodel="http://biomodels.net/model-qualifiers/" >
<rdf:Description rdf:about="#_180340"><dc:creator rdf:parseType="Resource">
<rdf:Bag><rdf:li rdf:parseType="Resource"><vCard:N rdf:parseType="Resource"><vCard:Family>Shapiro</vCard:Family>
<vCard:Given>Bruce</vCard:Given></vCard:N><vCard:EMAIL>bshapiro@jpl.nasa.gov</vCard:EMAIL>
<vCard:ORG><vCard:Orgname>NASA Jet Propulsion Laboratory</vCard:Orgname>
</vCard:ORG></rdf:li></rdf:Bag></dc:creator><dcterms:created rdf:parseType="Resource">
<dcterms:W3CDTF>2005-02-06T23:39:40</dcterms:W3CDTF></dcterms:created><dcterms:modified rdf:parseType="Resource">
<dcterms:W3CDTF>2006-11-14T21:55:41</dcterms:W3CDTF></dcterms:modified><bqmodel:is><rdf:Bag>

<rdf:li rdf:resource="http://www.ebi.ac.uk/biomodels/#BIOMD0000000003"/></rdf:Bag></bqmodel:is>
<bqmodel:isDescribedBy><rdf:Bag><rdf:li rdf:resource="http://www.pubmed.gov/#1833774"/></rdf:Bag>
</bqmodel:isDescribedBy><bqbiol:is><rdf:Bag>
<rdf:li rdf:resource="http://www.ncbi.nlm.nih.gov/Taxonomy/#8292"/></rdf:Bag></bqbiol:is>
<bqbiol:isVersionOf><rdf:Bag><rdf:li rdf:resource="http://www.geneontology.org/#GO:0000278"/>
<rdf:li rdf:resource="http://www.genome.jp/kegg/network/#hsa04110"/></rdf:Bag>
</bqbiol:isVersionOf><bqbiol:isHomologTo><rdf:Bag><rdf:li rdf:resource="http://www.reactome.org/#REACT_152"/>
</rdf:Bag></bqbiol:isHomologTo></rdf:Description></rdf:RDF></annotation>
    <listOfCompartments><compartment metaid="_230461" id="cell" name="cell" size="1" units="volume">
        <annotation><rdf:RDF xmlns:rdf="http://www.w3.org/1999/02/22-rdf-syntax-ns#" 
	xmlns:bqbiol="http://biomodels.net/biology-qualifiers/" 
	xmlns:bqmodel="http://biomodels.net/model-qualifiers/" >
<rdf:Description rdf:about="#_230461"><bqbiol:is><rdf:Bag>
<rdf:li rdf:resource="http://www.geneontology.org/#GO:0005623"/>
</rdf:Bag></bqbiol:is></rdf:Description></rdf:RDF></annotation></compartment></listOfCompartments>
    <listOfSpecies><species metaid="_230475" id="C" name="Cyclin" compartment="cell" 
	initialConcentration="0.01" substanceUnits="substance" spatialSizeUnits="volume">
        <annotation><rdf:RDF xmlns:rdf="http://www.w3.org/1999/02/22-rdf-syntax-ns#" 
	xmlns:bqbiol="http://biomodels.net/biology-qualifiers/" 
	xmlns:bqmodel="http://biomodels.net/model-qualifiers/" >
<rdf:Description rdf:about="#_230475"><bqbiol:isVersionOf><rdf:Bag>
<rdf:li rdf:resource="http://www.ebi.ac.uk/interpro/#IPR006670"/>
</rdf:Bag></bqbiol:isVersionOf></rdf:Description></rdf:RDF></annotation></species>
      <species metaid="_230495" id="M" name="CDC-2 Kinase" compartment="cell" 
	initialConcentration="0.01" substanceUnits="substance" spatialSizeUnits="volume">
        <annotation><rdf:RDF xmlns:rdf="http://www.w3.org/1999/02/22-rdf-syntax-ns#" 
	xmlns:bqbiol="http://biomodels.net/biology-qualifiers/" 
	xmlns:bqmodel="http://biomodels.net/model-qualifiers/" >
<rdf:Description rdf:about="#_230495"><bqbiol:hasVersion><rdf:Bag>
<rdf:li rdf:resource="http://www.uniprot.org/#P24033"/>
<rdf:li rdf:resource="http://www.uniprot.org/#P35567"/>
</rdf:Bag></bqbiol:hasVersion></rdf:Description></rdf:RDF></annotation></species>
      <species metaid="_230515" id="X" name="Cyclin Protease" compartment="cell" 
	initialConcentration="0.01" substanceUnits="substance" spatialSizeUnits="volume"/></listOfSpecies>
    <listOfParameters>
      <parameter id="V1" name="V1" constant="false"/><parameter id="V3" name="V3" constant="false"/>
      <parameter id="VM1" name="VM1" value="3"/><parameter id="VM3" name="VM3" value="1"/>
      <parameter id="Kc" name="Kc" value="0.5"/></listOfParameters>
    <listOfRules>
      <assignmentRule metaid="rule1" variable="V1"><math xmlns="http://www.w3.org/1998/Math/MathML">
          <apply><times/><ci> C </ci><ci> VM1 </ci><apply><power/><apply><plus/><ci> C </ci><ci> Kc </ci>
             </apply><cn type="integer"> -1 </cn></apply></apply></math></assignmentRule>
      <assignmentRule metaid="rule2" variable="V3"><math xmlns="http://www.w3.org/1998/Math/MathML">
          <apply><times/><ci> M </ci><ci> VM3 </ci></apply></math></assignmentRule></listOfRules>
    <listOfReactions><reaction metaid="_230535" id="reaction1" name="creation of cyclin" 
	reversible="false" fast="false">
        <annotation><rdf:RDF xmlns:rdf="http://www.w3.org/1999/02/22-rdf-syntax-ns#" 
	xmlns:bqbiol="http://biomodels.net/biology-qualifiers/" 
	xmlns:bqmodel="http://biomodels.net/model-qualifiers/" >
<rdf:Description rdf:about="#_230535"><bqbiol:isVersionOf><rdf:Bag>
<rdf:li rdf:resource="http://www.geneontology.org/#GO:0043037"/>
</rdf:Bag></bqbiol:isVersionOf></rdf:Description></rdf:RDF></annotation>
        <listOfProducts><speciesReference species="C"/></listOfProducts>
        <kineticLaw timeUnits="time" substanceUnits="substance">
          <math xmlns="http://www.w3.org/1998/Math/MathML">
            <apply><times/><ci> cell </ci><ci> vi </ci></apply></math>
          <listOfParameters><parameter id="vi" value="0.025"/></listOfParameters></kineticLaw></reaction>
      <reaction metaid="_230555" id="reaction2" name="default degradation of cyclin" 
	reversible="false" fast="false">
        <annotation><rdf:RDF xmlns:rdf="http://www.w3.org/1999/02/22-rdf-syntax-ns#" 
	xmlns:bqbiol="http://biomodels.net/biology-qualifiers/" 
	xmlns:bqmodel="http://biomodels.net/model-qualifiers/" >
<rdf:Description rdf:about="#_230555"><bqbiol:isVersionOf><rdf:Bag>
<rdf:li rdf:resource="http://www.geneontology.org/#GO:0008054"/>
</rdf:Bag></bqbiol:isVersionOf></rdf:Description></rdf:RDF></annotation>
        <listOfReactants><speciesReference species="C"/></listOfReactants>
        <kineticLaw timeUnits="time" substanceUnits="substance">
          <math xmlns="http://www.w3.org/1998/Math/MathML">
            <apply><times/><ci> C </ci><ci> cell </ci><ci> kd </ci></apply></math>
          <listOfParameters><parameter id="kd" value="0.01"/></listOfParameters></kineticLaw></reaction>
      <reaction metaid="_230575" id="reaction3" 
	name="cdc2 kinase triggered degration of cyclin" reversible="false" fast="false">
        <annotation><rdf:RDF xmlns:rdf="http://www.w3.org/1999/02/22-rdf-syntax-ns#" 
	xmlns:bqbiol="http://biomodels.net/biology-qualifiers/" 
	xmlns:bqmodel="http://biomodels.net/model-qualifiers/" >
<rdf:Description rdf:about="#_230575"><bqbiol:isVersionOf>
<rdf:Bag><rdf:li rdf:resource="http://www.geneontology.org/#GO:0008054"/>
</rdf:Bag></bqbiol:isVersionOf></rdf:Description></rdf:RDF></annotation>
        <listOfReactants><speciesReference species="C"/></listOfReactants>
        <listOfModifiers><modifierSpeciesReference species="X"/></listOfModifiers>
        <kineticLaw timeUnits="time" substanceUnits="substance">
          <math xmlns="http://www.w3.org/1998/Math/MathML">
            <apply><times/><ci> C </ci><ci> cell </ci><ci> vd </ci><ci> X </ci><apply>
                <power/><apply><plus/><ci> C </ci><ci> Kd </ci></apply>
		<cn type="integer"> -1 </cn></apply></apply></math>
          <listOfParameters><parameter id="vd" value="0.25"/>
            <parameter id="Kd" value="0.02"/></listOfParameters></kineticLaw></reaction>
      <reaction metaid="_230595" id="reaction4" name="activation of cdc2 kinase" 
	reversible="false" fast="false">
        <annotation><rdf:RDF xmlns:rdf="http://www.w3.org/1999/02/22-rdf-syntax-ns#" 
	xmlns:bqbiol="http://biomodels.net/biology-qualifiers/" 
	xmlns:bqmodel="http://biomodels.net/model-qualifiers/" >
<rdf:Description rdf:about="#_230595"><bqbiol:isVersionOf><rdf:Bag>
<rdf:li rdf:resource="http://www.ebi.ac.uk/IntEnz/#3.1.3.16"/>
<rdf:li rdf:resource="http://www.geneontology.org/#GO:0045737"/>
<rdf:li rdf:resource="http://www.geneontology.org/#GO:0006470"/>
</rdf:Bag></bqbiol:isVersionOf></rdf:Description></rdf:RDF></annotation>
        <listOfProducts><speciesReference species="M"/></listOfProducts>
        <kineticLaw timeUnits="time" substanceUnits="substance">
          <math xmlns="http://www.w3.org/1998/Math/MathML">
            <apply><times/><ci> cell </ci><apply><plus/><cn type="integer"> 1 </cn>
                <apply><times/><cn type="integer"> -1 </cn><ci> M </ci></apply></apply><ci> V1 </ci>
              <apply><power/><apply><plus/><ci> K1 </ci><apply><times/><cn type="integer"> -1 </cn>
		<ci> M </ci></apply><cn type="integer"> 1 </cn></apply><cn type="integer"> -1 </cn>
		</apply></apply></math>
          <listOfParameters><parameter id="K1" value="0.005"/></listOfParameters></kineticLaw></reaction>
      <reaction metaid="_230615" id="reaction5" name="deactivation of cdc2 kinase" 
	reversible="false" fast="false">
	<annotation><rdf:RDF xmlns:rdf="http://www.w3.org/1999/02/22-rdf-syntax-ns#" 
	xmlns:bqbiol="http://biomodels.net/biology-qualifiers/" 
	xmlns:bqmodel="http://biomodels.net/model-qualifiers/" >
<rdf:Description rdf:about="#_230615"><bqbiol:isVersionOf><rdf:Bag>
<rdf:li rdf:resource="http://www.ebi.ac.uk/IntEnz/#2.7.10.2"/>
<rdf:li rdf:resource="http://www.geneontology.org/#GO:0045736"/>
<rdf:li rdf:resource="http://www.geneontology.org/#GO:0006468"/>
</rdf:Bag></bqbiol:isVersionOf></rdf:Description></rdf:RDF></annotation>
        <listOfReactants><speciesReference species="M"/></listOfReactants>
        <kineticLaw timeUnits="time" substanceUnits="substance">
          <math xmlns="http://www.w3.org/1998/Math/MathML">
            <apply><times/><ci> cell </ci><ci> M </ci><ci> V2 </ci>
              <apply><power/><apply><plus/>
                  <ci> K2 </ci><ci> M </ci></apply><cn type="integer"> -1 </cn></apply></apply></math>
          <listOfParameters><parameter id="V2" value="1.5"/><parameter id="K2" value="0.005"/>
          </listOfParameters></kineticLaw></reaction>
      <reaction metaid="_230635" id="reaction6" name="activation of cyclin protease" 
	reversible="false" fast="false">
        <listOfProducts><speciesReference species="X"/></listOfProducts>
        <kineticLaw timeUnits="time" substanceUnits="substance">
          <math xmlns="http://www.w3.org/1998/Math/MathML">
            <apply><times/><ci> cell </ci><ci> V3 </ci>
              <apply><plus/><cn type="integer"> 1 </cn><apply>
                  <times/><cn type="integer"> -1 </cn><ci> X </ci></apply></apply>
              <apply><power/><apply><plus/>
                  <ci> K3 </ci><apply><times/><cn type="integer"> -1 </cn><ci> X </ci></apply>
                  <cn type="integer"> 1 </cn></apply><cn type="integer"> -1 </cn></apply></apply></math>
          <listOfParameters><parameter id="K3" value="0.005"/></listOfParameters></kineticLaw></reaction>
	<reaction metaid="_230655" id="reaction7" name="deactivation of cyclin protease" 
	reversible="false" fast="false">
        <listOfReactants><speciesReference species="X"/></listOfReactants>
        <kineticLaw timeUnits="time" substanceUnits="substance">
          <math xmlns="http://www.w3.org/1998/Math/MathML">
            <apply><times/><ci> cell </ci><ci> V4 </ci><ci> X </ci>
              <apply><power/><apply><plus/><ci> K4 </ci><ci> X </ci></apply>
                <cn type="integer"> -1 </cn></apply></apply></math>
          <listOfParameters><parameter id="K4" value="0.005"/>
            <parameter id="V4" value="0.5"/></listOfParameters></kineticLaw></reaction>
    </listOfReactions></model></sbml>


\end{verbatim} 

\end{document}